# Stochasticity, periodicity and coherent structures in partially mode-locked fibre lasers

D. Churkin[1,2,3,*], S. Sugavanam[1], N. Tarasov[1], S. Khorev[3,4], S.V. Smirnov[2], S.M. Kobtsev[2], and S.K. Turitsyn[1,2]

[1]Aston Institute of Photonic Technologies, Aston University, Birmingham, B4 7ET, United Kingdom,
[2]Novosibirsk State University, 630090, Novosibirsk, Russia.
[3]Institute of Automation and Electrometry SB RAS, 1 Ac. Koptyug Ave., Novosibirsk, 630090, Russia
[4]Zecotek Photonics, Inc., 1120-21331 Gordon Way, Richmond, BC V6W 1J9, Canada.

**Correspondence to**: Dmitry Churkin, e-mail: d.churkin@aston.ac.uk

**Physical systems with co-existence and interplay of processes featuring distinct spatio-temporal scales are found in various research areas ranging from studies of brain activity to astrophysics. Complexity of such systems makes their theoretical and experimental analysis technically and conceptually challenging. Here, we discover that radiation of partially mode-locked fibre lasers, while being stochastic and intermittent on short time scale, exhibits periodicity and long scale correlations over slow evolution from one round trip to another. The evolution mapping of intensity autocorrelation function allows us to reveal variety of spatio-temporal coherent structures and to experimentally study their symbiotic co-existence with stochastic radiation. Our measurements of interactions of noisy pulses over a time scale of thousands of non-linear lengths demonstrate that they have features of incoherent temporal solitons. Real-time measurements of spatio-temporal intensity dynamics are set to bring new insight into rich underlying nonlinear physics of practical active- and passive-cavity photonic systems.**

Understanding the fundamental physics behind pulse generation in non-linear fibre resonators is critically important for development of high-energy pulsed lasers. In conventional mode-locked lasers, the pulse energy is inversely proportional to the repetition rate, offering a straightforward way to boost up the energy of generated pulses by elongation of the laser cavity[1–3]. Fibre optics provides a unique opportunity to design laser systems with ultra-long cavities up to several hundred km long for linear-cavity Raman fibre lasers[4] and up to 25 km long for passively mode-locked lasers[2,5–6] reaching pulse energies in master oscillators (MO) of up to 4 μJ for 3 ns pulses.

However, long fibre systems are known for their complex non-linear behaviour where phase and amplitude noise arising from non-linear interaction, scattering, and other effects result in loss of coherence and stochastisation of radiation. Moreover, as the laser cavity is elongated, fully coherent lasing operation tends to grow more and more unstable, finally developing into stochastic generation with low coherence[3,7,8]. Noise-like[9] or double-scale femto-/pico-second noisy pulses have been observed in long-cavity partially mode-locked pulsed fibre lasers[10]. On the other hand, various dissipative coherent structures are generated in mode-locked lasers under certain conditions, such as soliton rains[11,12] and molecules[13], or dark solitons[14]. In addition, complex processes of their interaction like soliton explosions[15] and rogue wave generation[16–18] have also been observed. Most of the regimes mentioned so far were observed in in fibre lasers mode-locked due to non-linear polarisation evolution (NPE). In order to trigger mode locking in these lasers, polarisation controllers are used that allow conversion of input light of arbitrary polarisation to output light of any desired polarization. As it is well known, complete transformation of the radiation polarisation state requires several phase retarders with at least three degrees of freedom. Large number of parameters for setting the radiation polarisation state in combination with laser cavity properties and adjustable pump power gives rise to a plethora of pulsed generation regimes including both regimes with complete mode locking and those with partial mode locking, these latter featuring much worse pulse parameter stability compared to the former. However, there is no defined boundary between these regime types and it is possible to trigger any of them with different settings of the polarisation controllers, producing pulses of different structure even under complete mode locking[19] and different dynamics as far as optical turbulence[20,21]. In general, generation of coherent structures in complex nonlinear systems featuring stochastic dynamics and role of such objects in turbulent energy transfer are important fundamental science problems[22-23]. Demonstration of co-existence of short-scale coherent structures and stochastic radiation in photonic devices opens-up opportunities for detailed, high-accuracy experimental investigations of such complex nonlinear systems. Understanding and mastering this variety of non-linear generation regimes may lead to development of new laser types with new features and better performance.

   In theory, most of efforts to describe pulse generation in mode-locked lasers are focused on asymptotic regimes when coherent structure and/or stable pulses are already formed. Therefore, complex processes of interaction of coherent structures and stochastic radiation are usually missed out. In experiments, it is also a challenging task to identify different partially mode-locked regimes and to gain insight into complex pulse dynamics of mode-locked lasers and the impact of stochasticity on properties of noisy pulses. In spectral domain, dispersive Fourier transform[24] can be used to study stochastic-driven processes in passive systems[25] and lasers with noise-like pulses[7]. In time domain, averaged (over many pulses) measurements such as autocorrelation or FROG-based techniques[26] work well in case of stable coherent pulses. For stochastic radiation exhibiting complex dynamics, all differences between pulses resulting from noise will be cancelled out in measurements averaged over many of them. Single-shot techniques[27] and direct real-time intensity measurements are required to reveal internal stochastic structure of individual pulses.

   In this work, using an advanced experimental technique for real-time measurement of spatio-temporal intensity dynamics we discover internal periodicity in various stochastic generation regimes of fibre lasers partially mode-locked due to NPE and reveal a number of dissipative coherent structures including bright and dark ones generated within both stochastic pulses and inter-pulse background. We highlight interaction between stochastic pulses themselves over a time scale of thousands of non-linear lengths, showing similarity with incoherent temporal solitons. We believe that study of spatio-temporal regimes of partially mode-locked fibre lasers is not only relevant for better understanding of performance of such lasers, but it also offers unique experimental test-bed for high accuracy investigation of turbulent energy transfer, soliton gas and other fundamental non-linear science problems.

**Results**

**Stochasticity and periodicity in spatio-temporal regimes**

As a test-bed system for analysis of stochasticity and periodicity, we adopted a 1-km long normal-dispersion ring-cavity fibre laser, in which both complete and partial mode locking could be triggered due to NPE. The configuration is explained in detail in Supplementary Materials. We chose a relatively long cavity in order to have access to a variety of complex partially mode-locked regimes, which are more numerous in long NPE-based fiber lasers than completely mode-locked ones. In an arbitrary regime with partial mode locking, the laser generates long stochastically filled pulses, and despite being mode-locked and generating a well-resolved pulse train with fairly stable inter-pulse separation (see Fig. 1(a)), it may produce pulses of substantially different shape and intensity at different moments in time, as the measured intensity dynamics $I(t)$ reveals (Fig. 1). The experimental challenge here is to decide whether these radically differing temporal profiles belong to the same pulse breathing in the cavity under the influence of some complex non-linear processes leading to strong pulse re-shaping or the laser hops from one generation regime to another and thus cannot be considered stable.

To answer this critical question, we measure spatio-temporal intensity dynamics, $I(t, T)$, instead of analysing simple one-dimensional intensity dynamics, $I(t)$. Here the instantaneous fast time co-ordinate $t$ may be mapped (because the fastest process is linear propagation with the speed of light $c$) onto the real space co-ordinate along the fibre, $x$, via transformation $t - x/c$, while slow evolution time $T$ is measured in terms of number of cavity round-trips. For more details, see Supplementary Materials and Ref.[28]. By measuring how the instantaneous intensity pattern $I(t)$ evolves over many cavity round-trips, we immediately reveal internal periodicity of the pulse evolution (Fig. 1(c)) and complex dynamics accompanying its evolution. The pulse exhibits complex dynamics over slow evolution coordinate with a typical evolution time equivalent to hundreds of round-trips. The pulse has a shockwave-type trailing edge and a smooth leading front. Note that the laser operates in a stationary regime, and the slow evolution time $T$ is measured starting from the some arbitrary moment of time.

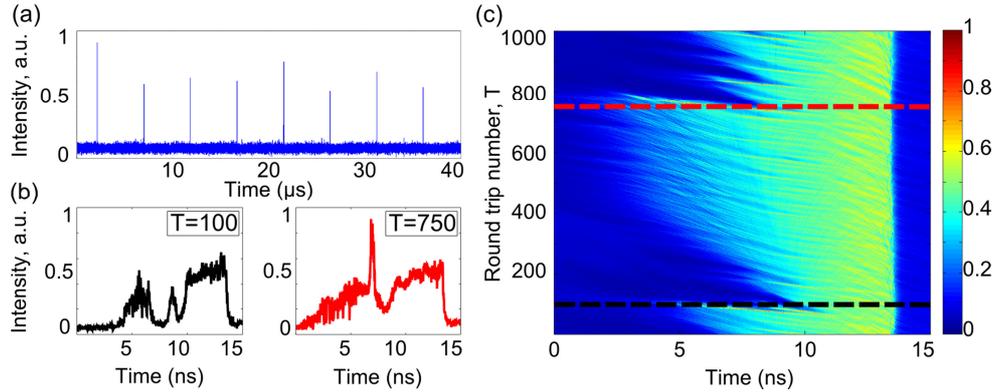

**Figure 1 | Temporal and spatio-temporal intensity dynamics of a partially mode-locked long fibre laser.** (a) Intensity dynamics I(t) demonstrating a noticeable variation in the intensity of the radiation from pulse to pulse. (b) Noise-like pulse has substantially different temporal shapes at different moments of time showing stochastic nature of the laser regime. (c) Corresponding spatio-temporal intensity dynamics plotted in a reference frame co-moving with a pulse reveals internal periodicity in the stochastic laser radiation. Intensity is colour-coded on a linear scale. The dotted line corresponds to the values of the evolution time at which intensity dynamics of panel (b) is plotted. Video S1 shows the pulse evolution.

Further on, we set the laser to generate different types of stochastic pulses with different spatio-temporal structure by adjusting the polarisation controller or the pump power, (Fig. 2). Although stochastic pulse shapes appear very similar in all presented regimes, these regimes are very different in their periodicity properties over the slow evolution time coordinate. They vary from one featuring completely homogenous and nearly stationary evolution over round-trips, Fig. 2(a), to regimes with well-pronounced evolution periodicity shown in Fig. 2(b, c), and, finally, to those where the pulse tends to be localised over both the spatial and evolution coordinates, as in Fig. 2(d). In all cases, except Fig. 2(a), the pulse shape is subject to pronounced modification in the course of its evolution. Fig. 2 clearly demonstrates that in the case of complex non-linear dynamics, generation should be characterised in terms of spatio-temporal intensity dynamics rather than as simple intensity dynamics over time.

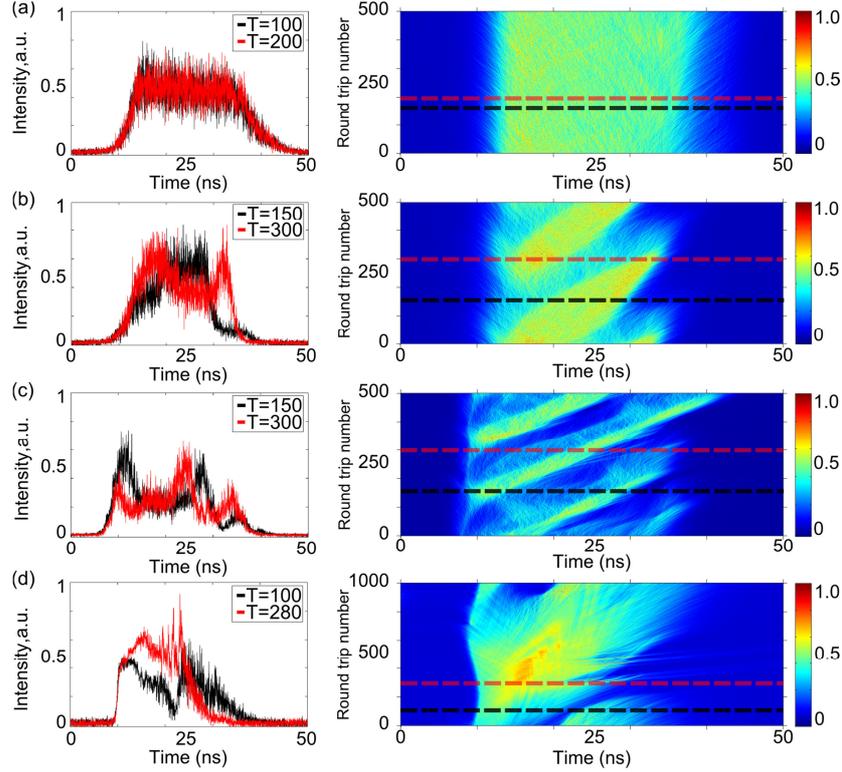

**Figure 2 | Different spatio-temporal regimes of noise-like pulses.** (a) A stochastic pulse stable over the slow evolution coordinate. (b, c) Examples of stochastic pulses having well-pronounced periodicity over the evolution coordinate. (d) A stochastic pulse localised both over fast time $t$ and slow evolution coordinate $T$. The right column in all panels demonstrates spatio-temporal intensity dynamics, $I(t,T)$. The left column shows the corresponding pulse profiles measured at evolution time T indicated with dotted lines. See corresponding videos of regimes shown in (c) and (d) in Videos S2–S5.

**Evolution map of the intensity autocorrelation function**

Different types of coherent and dissipative structures may be embedded in stochastic radiation. To reveal constituents of the studied radiation, we utilise an autocorrelation analysis. Further on, we analyse the spatio-temporal regime presented in Fig. 1. We measure the intensity autocorrelation function (ACF) defined as $K(\tau) = \langle I(t) \cdot I(t+\tau) \rangle_t$ directly with a real-time oscilloscope to have access to large detuning times $\tau$, which is crucial for identification of coherent structures within stochastic radiation. The intensity autocorrelation function measured over a large number of round-trips has a comb-like structure comprising a series of peaks separated by $T_{\mathrm{RT}} = nL/c$ because the intra-cavity radiation evolves quasi-periodically in the laser cavity (with an approximate period equal to round-trip time $T_{\mathrm{RT}}$). Here $L$ is the length of the ring laser cavity and $n$ is the average refractive index. Zero-order ACF peak resulting from two overlaid pulse train replicas appears at $\tau = 0$ and has two time scales: a narrow autocorrelation peak (of about 100-ps width) sitting on top of a wider background (about 10 ns wide), Fig. 3(a). In our case, the narrow ACF peak corresponds to the typical time scale of intensity fluctuations defining the stochastic nature of the pulse, while the broad pedestal reflects the average width of the noise-like pulse. Note that although we observe a double-scale ACF, it is different from the one previously studied in mode-locked lasers[9,10]. Indeed, in our case of partially mode-locking, stochastic pulses are much broader, so the smaller scale in the zero-order ACF for broad pulses corresponds to the wide pedestal of highly-stable pulses in papers[9,10].

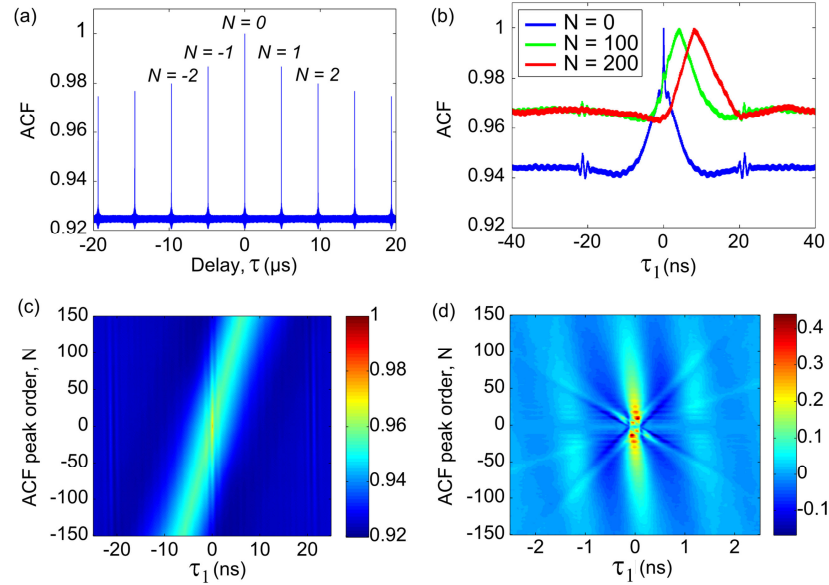

**Figure 3 | Revealing radiation constituents moving at different velocities via evolution analysis of the intensity auto-correlation function.** (a) Intensity ACF measured over large temporal detuning values. Peak separation equals the average cavity round-trip time. (b) Different-order peaks of intensity ACF (normalised to unity). (c, d) Autocorrelation function evolution map plotted in large (c) and small (d) scale over temporal detuning. Colour is used to show the amount of auto-correlation.

Each scale of the zero-order ACF peak means that there should be some typical structures of the corresponding temporal width within the total radiation. However, if different types of temporal structures have similar temporal width, they will be mixed up within the same scale of the zero-order ACF peak, becoming unresolvable there. Real-time measurements of intensity ACF over time delays $\tau$ up to hundreds of round-trip times $T_{RT}$ with an oscilloscope allow us to tell apart different types of radiation components even if they are of similar temporal width. To do that, we focus on the *N*-th order ACF peak located at time delay $\tau_N = N \cdot T_{RT}$. The *N*-th order ACF peak appears as a result of convolution of the intensity pattern with its replica shifted by N cavity round-trips (*i.e.* the same intensity pattern measured N cavity round-trips later). The radiation partials of different types may also have different group velocities resulting in different time needed for the studied structure to make a round-trip over the cavity. Thus, different structures having different group velocities may contribute to different sub-peaks superimposed on the main *N*-th order ACF peak. Indeed, higher-order intensity ACF peaks reveal certain signs of additional small amplitude peaks shifted off the centre of the main peak, Fig. 3(b).

Similar to time domain, one-dimensional representation of intensity ACF does not allow us to identify different structures rigorously. This is first of all caused by extremely small typical group velocity differences between different types of coherent structures and dispersive waves which could be as small as $10^{-4}$–$10^{-5}$, see Ref.[29]. Another obstacle is a small amount of energy in those structures compared to the total energy in the radiation leading to tiny, poorly resolved features in the $N$-th order ACF peak, Fig. 3(b), making any quantitative analysis impossible. To overcome this problem, we build a two-dimensional intensity ACF evolution map by plotting the ACF signal over time detuning $\tau_1$ and number $N$. Here $N = 0, \pm 1, \pm 2,$ *etc.* is the order of the ACF peak effectively representing the number of round-trips and time detuning $\tau_1 = \tau - N \cdot T_{\text{RT}}$ so that $|\tau_1| < T_{\text{RT}}/2$.

ACF evolution map provides a lot of information. For example, Fig. 3(c) demonstrates a set of vertical lines shifted by approximately 1 ns from the main pulse and further ones shifted by about ±21 ns. These features are linked to electrostriction effect in the optical fibre[30,31]. Being intensity-dependent, electrostriction is associated with the most energetic radiation component. There is also a broad slanted area corresponding to less energetic part of the stochastic radiation.

In general, any temporal structure with a constant group velocity different from that of the co-moving reference frame (used in the definition of $T_{\text{RT}}$) will give a straight slanted line on the ACF evolution map. We further zoom the ACF evolution map into the region of few-nanosecond time delays and subtract the contribution caused by stochastic radiation components (see Supplementary Materials for details). As a result, we find a number of small satellite auto-correlation peaks, whose position depends linearly on the order of correlation peaks, Fig. 3(d). Figure 3(d) is one of the central points of our work. This approach can be used for characterisation of coherent structures with varying parameters that maybe overlooked by other techniques when such structures are embedded into the stochastic radiation background. Different peaks correspond to structures of distinct types. The presence of different structures means, in particular, that the auto-correlation function in its zero-order has actually multiple overlapping scales. By measuring the angle of each satellite line, the group velocity difference of each structure may be determined to precision as high as $10^{-5}$–$10^{-6}$ and further used to adjust the value of $T_{\text{RT}}$ and to finally plot the spatio-temporal intensity dynamics, $I(t, T)$, in a co-moving reference frame selected so as to almost immobilise the structures of interest and, therefore, make them directly detectable.

**Coherent structures within the stochastic radiation**

To further explore distinct parts of radiation propagating at different velocities, we use various values of group velocities obtained from ACF evolution map. First, we take the group velocity value differing by $4 \cdot 10^{-5}$ from the initial value. We find that this corresponds to the inter-pulse background, which has low, but non-zero intensity and is supposed to contain pure noise as indicates simple intensity dynamics, Fig. S6. The presence of inter-pulse background is one of the unique features of partially mode-locked regimes, since inter-pulse background has hitherto never been observed in completely mode-locked regimes, not even those realised in ultra-long mode-locked fibre lasers[2], in which the pulse-to-pulse period approaches the upper-state lifetime of the active medium. Quite to the contrary, in partially mode-locking regimes, the spatio-temporal dynamics of the inter-pulse background plotted in an appropriate co-moving reference frame exhibits a prominent structure, Fig. 4(a), corresponding to the laminar state recently observed in Raman fibre lasers[28].

Spatio-temporal representation of laser dynamics allows us to study temporal properties of the radiation separately for different clusters of radiation and get better understanding of their internal structure. For an example, here we filter out the main stochastic pulse from the laminar inter-pulse background and study intensity ACF and probability density function (PDF) measured for the inter-pulse background only, Fig. 4(b, c). The ACF level is close to unity and PDF has a prominent bell-shaped peak that proves the coherent nature of the background. Existence of stochastic pulses propagating over a laminar background makes partially mode-locked fibre lasers very similar to classical hydrodynamic pipe systems with turbulent puffs propagating within the laminar fluid flow[32]. Thus, classical questions on how critical parameters govern the turbulence onset and decay can be also posed in fibre-laser experiments using partially mode-locked lasers.

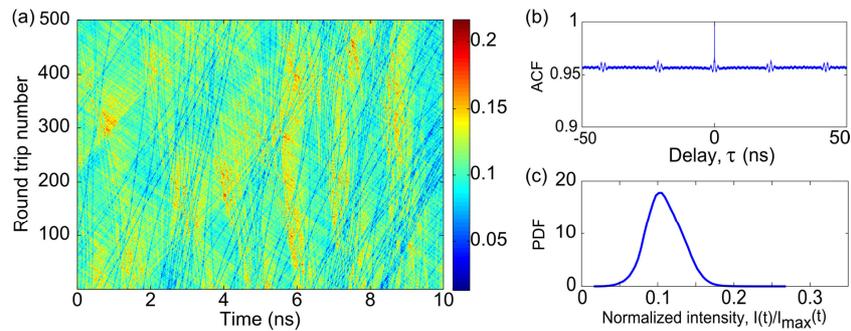

**Figure 4 | Laminar inter-pulse background supporting propagation of dark/grey solitons.** (a) Spatio-temporal intensity dynamics of the inter-pulse radiation. Laminar inter-pulse background is coherent with suppressed intensity fluctuations. The background is furrowed by numerous dark and grey solitons (blue curved traces) living on the background. (b) Intensity autocorrelation function and (c) intensity probability density function measured for the inter-pulse background only (excluding main pulses). The intensity is normalised to the peak value occurring in the regime.

Note dark traces on the inter-pulse background, Fig. 4(a). Each trace has a typical length of hundreds of non-linear lengths, therefore they must correspond to coherent structures. It is known that dark and grey solitons may emerge on a condensate (laminar) background and be stable there[33]. Dark and grey solitons also proved to be generated on a condensate background in laminar regime of quasi-CW Raman fibre lasers[28]. In[34], it was numerically found that dark/grey solitons are generated at the initial stage of radiation build-up in passively mode-locked fibre lasers, although they eventually decay.

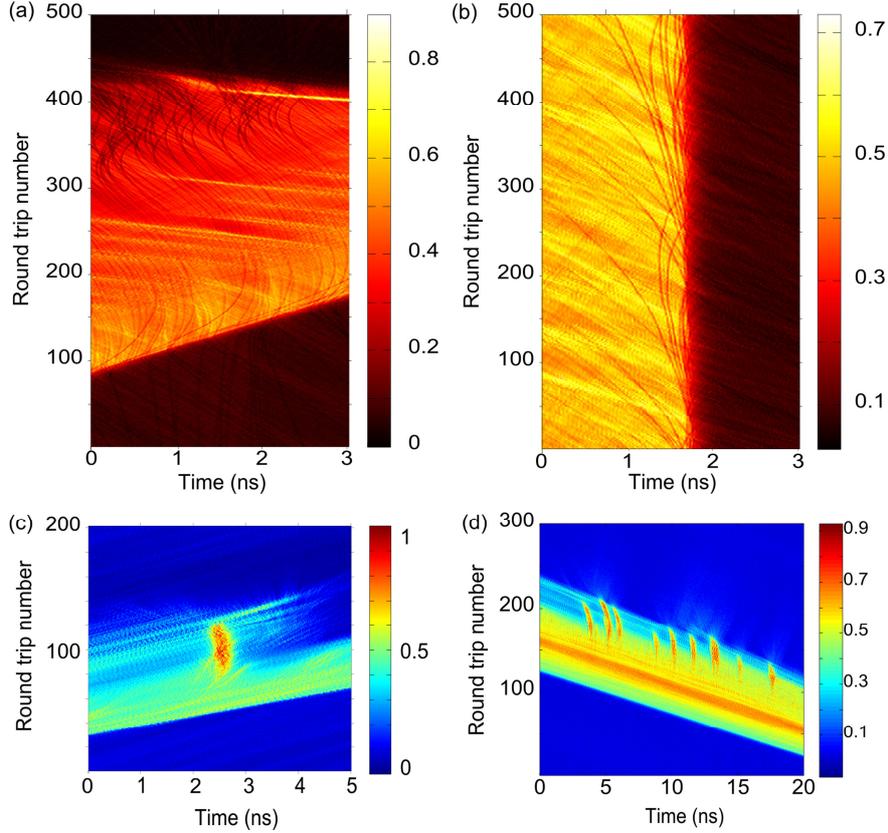

**Figure 5 | Co-existence and interaction of dark and bright coherent structures with a stochastic pulse.** (a, b) Interaction of dark and grey solitons with a stochastic pulse. (c, d) Generation of bright coherent structures within a stochastic pulse. Different speeds of the co-moving reference frame are used. Panels (a–c) present spatio-temporal dynamics for the regime of Fig. 1. Panel (d) corresponds to the spatio-temporal regime shown in Fig. 2(d).

Dark and grey solitons propagate on condensate/laminar background at different speeds depending on their amplitude, and interact non-linearly between each other (interaction is visually represented as curved dark traces on the spatio-temporal intensity dynamics, Fig. 4(a), which is a sign of changing speed of these structures). Choosing an appropriate co-moving reference frame speed, which we determine from the intensity ACF evolution map of Fig. 3(d), we can highlight intriguing interaction of dark/grey solitons with a stochastic pulse. There are different scenarios of interaction depending on the velocity difference between solitons and main pulse, Fig. 5(a, b). Some dark solitons, after hitting the kink-type edge of the pulse, enter the pulse, become trapped inside, acquiring the pulse's speed, and eventually leave the pulse through its other edge, assuming a speed almost equal to the initial one, Fig. 5(a) (see also Fig. S7 for details). Other solitons "bounce" back from the pulse as demonstrated in Fig. 5b. This effect resembles a full internal reflection in geometrical optics.

In general, we have found dark and grey solitons in every regime shown in Fig. 2. However, details of dark/grey soliton interaction with a pulse may be different in each particular case. Note that mutual dynamics of dark and grey solitons, which enter noise-like pulses, propagate in them, and eventually exit, partially defines the temporal pulse shape making it appear more stochastic in one-dimensional time domain data representation $I(t)$.

Apart from dark and grey solitons, we also observe simultaneous generation of bright coherent structures. Indeed, the spatio-temporal dynamics presented in Fig. 1(d) and further on in Fig. 5(a) in a different moving frame of reference reveals some short high-intensity events propagating at a speed different from that of the pulse. In a reference frame moving with the pulse, such events look like high-intensity extreme walls, see Fig. S8. Spatio-temporal dynamics plotted using a corresponding reference frame co-moving with bright structures allows us to visualize such bright structures and estimate their life-time over the slow evolution time, Fig. 6(c). These stochastic bursts of light persist for up to 30 round-trips over the evolution coordinate (which corresponds to propagation distance of up to 30 km), therefore they must be coherent. Note that as they travel at a speed substantially different from that of the main pulse, they propagate freely through the large-scale pulse envelope. It seems that generation of bright coherent structures within stochastic pulses is a general property of the system, as we find them in many regimes. For an example, see Fig. 5(d) where quasi-periodic bright coherent structures are highlighted. There, the same regime is presented as in Fig. 2(d).

Finally, one can examine extremely long spatio-temporal dynamics over tens of thousands of round-trips at the expense of fast time resolution. It has been found that noise-like pulses may be stable in the long-term, Fig. 6(a). Long-term evolution of the inter-pulse background (having laminar nature on the short time scale) reveals existence of low intensity spatio-temporal waves (see Fig. 6(a), colour inset; note different colour code used to highlight low intensity waves). Interestingly, stochastic pulses localised both over the fast time and slow evolution coordinate presented in Fig. 2(d) also exhibit high long-term stability and periodicity, as demonstrated in Fig. 6(b). The period over the evolution coordinate is around 400 round-trips (approximately 2 ms in terms of evolution time). Further adjusting the pump power and polarisation controllers, we observe another example of localized noise-like pulses, Fig. 6(c). In this case, the pulses are about 10 ns wide and survive for longer than 100 round-trips over the evolution coordinate. More interestingly, in this regime stochastic pulses actually interact with each other on the evolution time scale of thousand round-trips, which process manifests itself as scattering of one stochastic pulse upon another. In these scattering events, energy can be transferred from one pulse to another, further leading to emergence of new pulses. Interacting pulses also change their propagation speed. Overall, the interaction looks similar to interaction of coherent structures.

Lastly, the laser demonstrates yet another kind of non-trivial breathing spatio-temporal evolution. Namely, a train of stochastic pulses organised in an interacting ensemble can be generated, Fig. 6(d). In this regime, generated pulses transfer energy from the first pulse in the train to the following ones, so that eventually the first pulse is exhausted, and a new pulse at the trailing edge of the train is born (see Video S9).

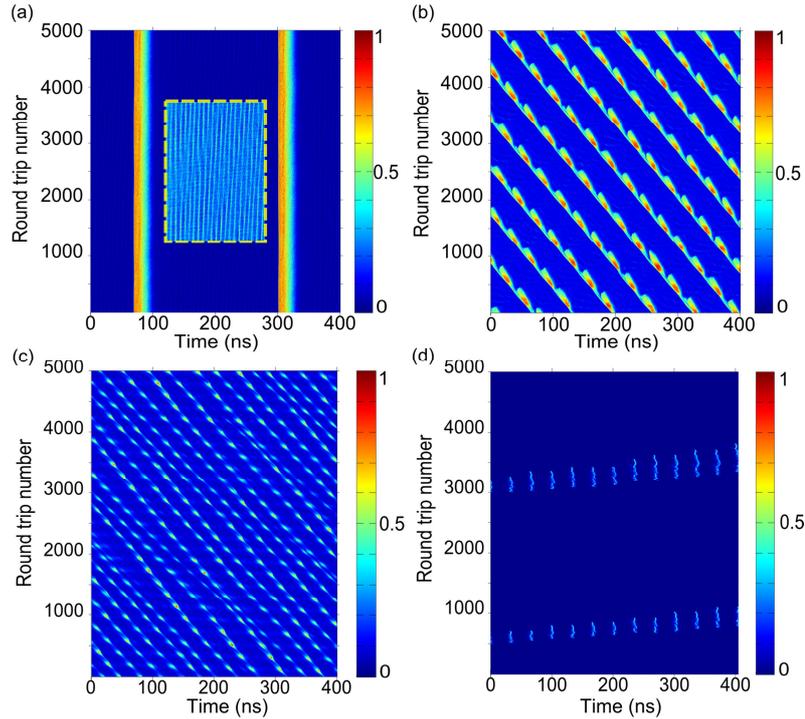

**Figure 6 | Long-term periodicity and interaction of noise-like pulses.** (a) Generation of stochastic pulses stable over slow evolution time. The inset uses a different colour code to highlight low-intensity inter-pulse dynamics. (b) Non-interacting and (c) interacting localised noise-like pulses. (d) Generation of a train of noise-like pulses. See Videos S6-S9.

**Discussion**

The presented experimental results greatly expand the capabilities of experimental characterisation of pulsed fibre laser systems and impart new insights into extremely complex and diverse nonlinear physics behind their operation. Complex and visibly stochastic generation regimes are found to have internal periodicity over slow evolution time, so these regimes may be experimentally classified on the basis of their spatio-temporal properties rather than temporal properties alone. This is of particular importance as long-range interactions and correlations over evolution time in fibre lasers could be missed out in conventional numerical simulations because of numerical restrictions. The proposed evolution mapping technique for intensity auto-correlation function reveals co-existence of coherent structures of various temporal scales and stochastic dynamics in partially mode-locked fibre lasers that may be undetectable by conventional measurement methods. Additional insight can be obtained when real-time measurements of intensity spatio-temporal dynamics are combined with spectral filtering techniques to associate specific spectral components with different spatio-temporal structures. This approach paves the way to experimental studies of mode correlations build-up from the initial noise at the start of mode-locked operation. More generally, any light source exhibiting complex intermittent temporal dynamics may also have hidden long-term correlations, which lead to different spatio-temporal dynamics and can be now experimentally studied.

We anticipate that experimental characterisation of spatio-temporal dynamics of different types of systems (not limited to cavity- and fibre-based ones) will result in deeper understanding of underlying complex physical processes and may further result in development of new laser types. In particular, presence of a number of coherent structures embedded into stochastic radiation may define properties of wave turbulent energy transfer and spectral performance of such lasers.

**Acknowledgments**
We thank Prof. L. Melnikov, Prof. C. Conti, Dr. A. Picozzi and Dr. S. Skipetrov for fruitful discussions, Dr. A. Ivanenko for technical assistance and OFS Denmark for fibres provided. This work was supported by the European Research Council (project ULTRALASER), the Russian Science Foundation (Grant No. 14-21-00110), Russian Ministry of Science and Education (agreements No. 14.B25.31.0003 and No. 3.162.2014/K) and Presidential grant for young researches (Russia).

**Author Contributions**
S.V.K. and S.M.K. designed the laser. S.S. conceived the idea of the ACF evolution map and made measurements. S.S. and N.T. processed the data. All authors analysed the data. D.V.C. and S.K.T. wrote the paper. D.V.C. conceived and supervised the project.

**Competing financial interests**
The authors declare no competing financial interests.

**References**

1. W. H. Renninger and F. W. Wise, Dissipative soliton fiber lasers (Fiber lasers, O. G. Okhotnikov (Ed.), Wiley-VCH Verlag GmbH & Co. KgaA, 2012), Chap. 4.
2. S. V. Smirnov, S. M. Kobtsev, S. V. Kukarin, and S. K. Turitsyn, Mode-locked fibre lasers with high-energy pulses (Laser Systems for Applications, K.Jakubczak (Ed.), InTech, 2011), Chap. 3.
3. P. Grelu and N. Akhmediev, "Dissipative solitons for mode-locked lasers," Nat. Photonics **26**, 84–92 (2012).
4. S.K. Turitsyn, J.D. Ania-Castañón, S.A. Babin, V. Karalekas, P. Harper, D. Churkin, S.I. Kablukov, A.E. El-Taher, E.V. Podivilov, and V.K. Mezentsev, "270-km ultralong Raman fiber laser,". Phys. Rev. Lett. **103**, 133901 (2009).
5. S. Kobtsev, S. Kukarin and Yu. Fedotov, " Ultra-low repetition rate mode-locked fiber laser with high-energy pulses," Opt. Express **16**, 21936-21941 (2008).
6. B.N. Nyushkov, A.V. Ivanenko, S.M. Kobtsev, S.K. Turitsyn, C. Mou, L. Zhang, V.I. Denisov, V.S. Pivtsov, "Gamma-shaped long-cavity normal-dispersion modelocked Er- laser for sub-nanosecond high-energy pulsed generation," Las. Phys. Lett. **9**, 59-67 (2012).

7. A.F. Runge, C. Aguergaray, N.G. Broderick, and M. Erkintalo, "Coherence and shot-to-shot spectral fluctuations in noise-like ultrafast fiber lasers," Opt. Lett. **38**, 4327-4330 (2013).
8. S.M. Kobtsev and S.V. Smirnov, "Fiber lasers mode-locked due to nonlinear polarization evolution: golden mean of cavity length," Las. Phys. **21**, 272-276 (2011).
9. M. Horowitz, Y. Barad, and Y. Silberberg, "Noiselike pulses with a broadband spectrum generated from an erbium-doped fiber laser," Opt. Lett. **22**, 799-801 (1997).
10. S. Kobtsev, S. Kukarin, S. Smirnov, S. Turitsyn, and A. Latkin, "Generation of double-scale femto/pico-second optical lumps in mode-locked fiber lasers," Opt. Express **23**, 20707-20713 (2009).
11. S. Chouli and P. Grelu, "Rains of solitons in a fiber laser," Opt. Express **17**, 11776-11781 (2009).
12. S. Chouli and P. Grelu, "Soliton rains in a fiber laser: An experimental study," Physical Review A, **81**, 063829 (2010).
13. Stratmann, M., Pagel, T. & Mitschke, F. Experimental observation of temporal soliton molecules. Phys. Rev. Lett. 95, 143902 (2005).
14. D. Tang, J. Guo, Y. Song, H. Zhang, L. Zhao, and D. Shen, "Dark soliton fiber lasers," Opt. Express **22**, 19831-19837 (2014).
15. S. T. Cundiff, J.M. Soto-Crespo, and N. Akhmediev, "Experimental evidence for soliton explosions," Physical Review Letters, **88**, 073903 (2002).
16. C. Lecaplain, P. Grelu, J.M. Soto-Crespo, and N. Akhmediev, "Dissipative rogue waves generated by chaotic pulse bunching in a mode-locked laser," Physical review letters, **108**, 233901 (2012).
17. A.F.J. Runge, C. Aguergaray, N.G.R. Broderick, and M. Erkintalo, "Raman rogue waves in a partially mode-locked fiber laser," Optics letters **39**, 319-322 (2014).
18. C. Lecaplain and P. Grelu, "Rogue waves among noiselike-pulse laser emission: An experimental investigation," Physical Review A **90**, 013805 (2014).
19. S. Smirnov, S. Kobtsev, S. Kukarin, and A. Ivanenko, "Three key regimes of single pulse generation per round trip of all-normal-dispersion fiber lasers mode-locked with nonlinear polarization rotation," Opt. Express **20**, 27447-27453 (2012).
20. F. X. Kärtner, D. M. Zumbühl, and N. Matuschek, "Turbulence in mode-locked lasers," Phys. Rev. Lett. **82** (22), 4428–4431 (1999).
21. S. Wabnitz, "Optical turbulence in fiber lasers," Opt. Lett. **39**, 1362-1365 (2014).
22. B. Rumpf, A.C. Newell, and V.E. Zakharov, V. E. Turbulent transfer of energy by radiating pulses. Physical review letters, 103(7), 074502 (2009).
23. V.E. Zakharov, A.N. Pushkarev, V.F. Shvetz, and V.V. Yan'kov, *Solitonic turbulence*, JETP Lett. **48**, 83-85 (1988).
24. K. Goda and B. Jalali, "Dispersive Fourier transformation for fast continuous single-shot measurements," Nat. Photonics, **7**, 102-112 (2013).

25. D. R. Solli, G. Herink, B. Jalali, and C. Ropers, "Fluctuations and correlations in modulation instability," Nat. Photonics, **6**, 463-468 (2012).
26. R. Trebino, *Frequency-Resolved Optical Gating: The Measurement of Ultrashort Laser Pulses: The Measurement of Ultrashort Laser Pulses, Vol. 1* (Springer, 2000).
27. T.C. Wong, M. Rhodes, and R. Trebino, "Single-shot measurement of the complete temporal intensity and phase of supercontinuum," Optica **1**, 119-124 (2014).
28. E.G. Turitsyna, S.V. Smirnov, S. Sugavanam, N. Tarasov, X. Shu, S.A. Babin, E.V. Podivilov, D.V. Churkin, and S.K. Turitsyn, "The laminar-turbulent transition in a fibre laser," Nat. Photonics **7**, 783-786 (2013).
29. Akhmediev N., Ankiewicz A. (eds). Dissipative Solitons: From Optics to Biology and Medicine Springer (2008).
30. E.M. Dianov, A.V. Luchnikov, A.N. Pilipetskii, and A.M. Prokhorov, "Long-range interaction of picosecond solitons through excitation of acoustic waves in optical fibers," Appl. Phys. B **54**, 175–180 (1992).
31. J. K. Jang, M. Erkintalo, S.G. Murdoch, and S. Coen, "Ultraweak long-range interactions of solitons observed over astronomical distances," Nat. Photonics **7**, 657-663 (2013).
32. K. Avila, D. Moxey, ., A. de Lozar, M. Avila, D. Barkley, and B. Hof, "The onset of turbulence in pipe flow," Science, **333**, 192-196 (2011).
33. Y.S. Kivshar and B. Luther-Davies, "Dark optical solitons: physics and applications," Physics Reports **298**, 81-197 (1998).
34. E.J.R. Kelleher and J.C. Travers, "Chirped pulse formation dynamics in ultra-long mode-locked fiber lasers," Opt. Lett., 39, 1398-1401 (2014).